\documentclass[twocolumn,showpacs]{revtex4}
\usepackage{graphicx}
\usepackage{dcolumn}
\usepackage{bm}

\def\stacksymbols #1#2#3#4{\def\theguybelow{#2}
    \def\verticalposition{\lower#3pt}
    \def\spacingwithinsymbol{\baselineskip0pt\lineskip#4pt}
    \mathrel{\mathpalette\intermediary#1}}
\def\intermediary#1#2{\verticalposition\vbox{\spacingwithinsymbol
    \everycr={}\tabskip0pt
    \halign{$\mathsurround0pt#1\hfil##\hfil$\crcr#2\crcr
        \theguybelow\crcr}}}

\begin{document}

\title{Criteria for the experimental observation of
multi-dimensional optical solitons in saturable media}
\author{Yi-Fan Chen}
\email[Corresponding author: ]{yc245@cornell.edu}
\affiliation{Department of Applied and Engineering Physics, Cornell University, 212 Clark
Hall, Ithaca, NY 14853, USA}
\author{Kale Beckwitt}
\affiliation{Department of Applied and Engineering Physics, Cornell University, 212 Clark
Hall, Ithaca, NY 14853, USA}
\author{Frank W. Wise}
\affiliation{Department of Applied and Engineering Physics, Cornell University, 212 Clark
Hall, Ithaca, NY 14853, USA}
\author{Boris A. Malomed}
\affiliation{Department of Interdisciplinary Studies, Faculty of Engineering, Tel Aviv
University, Tel Aviv 66978, Israel}
\date{\today}

\begin{abstract}
Criteria for experimental observation of multi-dimensional optical
solitons in media with saturable refractive nonlinearities are
developed. The criteria are applied to actual material parameters
(characterizing the cubic self-focusing and quintic
self-defocusing nonlinearities, two-photon loss, and
optical-damage threshold) for various glasses. This way, we
identify operation windows for soliton formation in these glasses.
It is found that two-photon absorption sets stringent limits on
the windows. We conclude that, while a well-defined window of
parameters exists for two-dimensional solitons (spatial or
spatiotemporal), for their three-dimensional spatiotemporal
counterparts such a window \emph{does not} exist, due to the
nonlinear loss in glasses.
\end{abstract}

\maketitle

\section{Introduction}

Solitons are localized wave packets and/or beams that result from
the balance of the linear and nonlinear responses of a physical
system. Depending on the physical properties of the underlying
system, solitons take different forms. They can be hydrodynamic
wave packets, such as solitary waves in the ocean \cite{solocean}
and atmosphere \cite{solatom}. They can also be spin-wave packets,
such as magnetic solitons \cite{spinsol1,spinsol2}. Bose-Einstein
condensates provide a medium to produce matter-wave solitons
\cite{becsol}. Other examples of soliton dynamics can be found in
a wide variety of fields, including astrophysics, plasma physics,
nuclear physics, and even metabolic biology \cite{astro1, astro2,
nucle1, bio1}, among others. Very accurate experiments have been
performed with topological solitons (fluxons) in long Josephson
junctions, including a recent direct observation of their
macroscopic quantum properties \cite{Ustinov}.

Solitons in optics, which are known in their temporal, spatial,
and spatiotemporal varieties (the latter ones being frequently
called \textquotedblleft light bullets\textquotedblright ),
constitute, perhaps, the most versatile and well-studied (both
theoretically and experimentally) class of solitons in physics. In
particular, temporal solitons in optical fibers \cite{agrawal}
have recently made a commercial debut in high-speed
telecommunications links \cite{agrawal,Australia}. It has been
pointed out that multi-dimensional (multi-D) spatiotemporal
optical solitons can be used in the design of high-speed
all-optical logic gates and, eventually, in all-optical
computation and communications systems \cite{soliton_gate}.

The balance of linear and nonlinear dynamical features is only the
first step in the soliton formation. Securing the stability of
this balance is the second, equally important step. A well-known
difficulty is that the most common optical nonlinearity -- the
Kerr effect in dielectrics -- gives rise to soliton solutions
which are unstable in more than one dimension against the wave
collapse, as discussed (in particular) in original papers
\cite{jetpcollapse, soliton_collapse4} and in the review
\cite{Luc}. Several mechanisms that can suppress the collapse have
been investigated. These include saturation of the Kerr
nonlinearity \cite {satkerr}, higher-order dispersion or
diffraction (also referred to as "non-paraxiality")
\cite{non_parax}, multiphoton ionization \cite {multi-photon}, and
stimulated Raman scattering (SRS) \cite{srs1,srs2}. In particular,
importance of the multi-photon absorption and SRS for the
spatiotemporal self-focusing of light in the Kerr medium was
inferred from experimental data in Ref. \cite{Shimshon}. However,
these mechanisms eventually reduce the intensity and cause the
pulse to expand in time and space, precluding the achievement of
multi-dimensional solitons \cite{frank}.

Different versions of the saturable nonlinearity (which implies
saturation of the cubic nonlinear susceptibility, $\chi ^{(3)}$,
in high-intensity fields) have been studied theoretically in
detail. It was shown that both rational \cite{yc, enns_1,
edmundson_1, edmundson_2, blair} and cubic-quintic (CQ)
\cite{quiroga,Anton,boris} variants of the saturation readily
support stable two-dimensional (2D) and three-dimensional (3D)
solitons. A difference between them is that the former cannot
stabilize \textquotedblleft spinning\textquotedblright\ solitons
with an intrinsic vorticity, but the CQ nonlinearity makes it
possible, in the 2D \cite {Spain,Lucian,BobPego} and even 3D
\cite{spinning-bullet} cases.

The first observation of a self-trapped beam in a Kerr medium was
reported by Bjorkholm and Ashkin in 1974 \cite{bj}. The experiment
was done in sodium vapor around the $D_{2}$ transition line, and
self-focusing arose from strong saturation of the transition (i.e.
saturation of the linear susceptibility, $\chi ^{(1)}$). Studies
of 2D solitons have made rapid progress since the mid-1990's in
the study of two new nonlinearities featuring saturation. Segev
\textit{et al}. predicted that the photorefractive (PR) effect in
electro-optic materials could be exploited to create an effective
saturable nonlinear index of refraction that would support
solitons \cite{segevprl}. PR solitons were observed experimentally
soon afterward \cite{duree}. In parallel to this, there was a
resurgence of interest in the so-called cascading nonlinearity,
which is produced by the interaction of two or three waves in
media with quadratic ($\chi ^{(2)}$) nonlinear susceptibility.
Both 1D and multi-D solitons in the quadratic media had been
studied theoretically in numerous works (see reviews \cite {Jena}
and \cite{buryak}). Stationary 2D spatial solitons (in the form of
self-supporting cylindrical beams) were first generated in
quadratic media by Torruellas \textit{et al}. \cite{torru}. Later,
Di Trapani \textit{et al}. observed temporal $\chi ^{(2)}$
solitons \cite{tquadratic}, and, finally, spatiotemporal solitons
were produced by Liu \textit{et al}. \cite {quadratic_sts,
quadratic_sts2}. Under appropriate conditions, both the PR and
cascading nonlinearities may be modeled as saturable
generalizations of the Kerr nonlinearity (despite the fact that
the PR media are, strictly speaking, non-instantaneous, nonlocal,
and anisotropic). However, to date, multi-D solitons in true
saturable Kerr media have not been observed.

In this work, we examine the possibility of stabilizing solitons
(arresting the collapse) in saturable Kerr media \cite{satkerr},
from the perspective of experimental implementation. Existing
theories provide for parameter regions where formation of stable
solitons is possible, but neglect linear and nonlinear losses, as
well as other limitations, such as optical damage in
high-intensity fields \cite{footnote}. First, we propose a
criterion for acceptable losses, and determine the consequences of
the loss for the observation of soliton-like beams and/or pulses.

Then, as benchmark saturable Kerr media, we consider nonlinear
glasses. Direct experimental measurements of the higher-order
nonlinearities and nonlinear (two-photon) loss  in a series of
glasses allow us to link the theoretical predictions to
experimentally relevant values of the parameters. As a result, we
produce \textquotedblleft maps\textquotedblright\ of the
experimental-parameter space where 2D and 3D solitons can be
produced. To our knowledge, this is the first systematic analysis
of the effects of nonlinear absorption on soliton formation in
saturable Kerr media. We conclude that it should be possible,
although challenging, to experimentally produce 2D spatial and 2D
spatiotemporal solitons in homogeneous saturable media.
Spatiotemporal solitons require anomalous group-velocity
dispersion (GVD). Under conditions relevant to saturation of the
Kerr nonlinearity, material dispersion is likely to be normal. In
that case, anomalous GVD might be obtained by pulse-tilting e.g.
On the other hand, the prospects for stabilizing 3D solitons seem
poor, even ignoring the need for anomalous GVD. This conclusion
suggests that qualitatively different nonlinearities, such as
$\chi ^{(2)}$, may be more relevant to making light bullets.

We focus on Gaussian beam profiles, which are prototypical
localized solutions. Very recent work has shown that nonlinear
loss can induce a transition from Gaussian to conical waves, which
can be stationary and localized \cite{filinochannel,arxiv}. The
conical waves are very interesting, but represent a different
regime of wave propagation from that considered here.

The rest of the paper is organized as follows. The theoretical
analysis of the necessary conditions for the formation of the 2D
and 3D solitons is presented in Section 2. Results of experimental
measurements of the nonlinear parameters (cubic and quintic
susceptibilities, and two-photon loss) in a range of glasses are
reported in Section 3. Final results, in the form of windows in
the space of physical parameters where the solitons may be
generated in the experiment, are displayed in Section 4, and the
paper is concluded by Section 5.

\section{Theoretical analysis of necessary conditions for the existence of
two- and three-dimensional solitons: lossless systems}

Evolution of the amplitude $E$ of the electromagnetic wave in a lossless
Kerr-like medium with anomalous GVD obeys the well-known scaled equation
\cite{enns_1,edmundson_1,edmundson_2,blair}
\begin{equation}
iE_{z}+\frac{1}{2}(E_{xx}+E_{yy}+E_{tt})+f(I)E=0\ ,  \label{waveeq}
\end{equation}
where $z$ and $\left( x,y\right) $ are the propagation and
transverse coordinates, and $t$ is the reduced temporal variable,
and $f(I)$ is proportional to the nonlinear correction to the
refractive index $\Delta n(I) $. In the Kerr medium proper, the
refractive index is $n(I)\equiv n_{0}+\Delta n(I)=n_{0}+n_{2}I$,
which, as was mentioned above, gives rise to unstable multi-D
solitons, including the weakly unstable \textit{ Townes soliton}
in the 2D case \cite{Luc}. Upon the propagation, the unstable
solitons will either spread out or collapse towards a singularity,
depending on small perturbations added to the exact soliton
solution.

Conditions for the soliton formation are usually expressed in
terms of the normalized energy content, but from an experimental
point of view it is more relevant to express the conditions in
terms of intensity and size (temporal duration and/or transverse
width) of the pulse/beam. They can also be converted into the
dispersion and diffraction lengths, which are characteristics of
the linear propagation. We transform the results of Ref.
\cite{soliton_collapse4} to estimate the parameters of the 2D and
3D solitons in physical units. The transformation is based on the
fact that the solutions are scalable with the beam size. Without
losing generality, the estimation also assumes a Gaussian profile
for the solutions. The relations between the critical peak
intensity necessary for the formation of the soliton and
diffraction length, in SI units are:
\begin{equation}
I_{\mathrm{critical}}\approx \left\{
\begin{array}{r@{\quad \ \quad}l}
0.52\left( \frac{{n_{0}^{2}}}{n_{2}}\right) \left( \frac{\lambda
_{0}}{L_{
\mathrm{diffr}}}\right)  & \mathrm{for\ \ 2D}, \\
&  \\
0.79\left( \frac{{n_{0}^{2}}}{n_{2}}\right) \left( \frac{\lambda
_{0}}{L_{ \mathrm{diffr}}}\right)  & \mathrm{for\ \ 3D},
\end{array}
\right.   \label{critical}
\end{equation}
where $L_{\mathrm{diffr}}={2\pi n_{0}w_{0}^{2}}/{\lambda _{0}}$ is
the diffraction length of the beam with the waist width $w_{0}$.
Eq. (\ref{critical}) is easy to understand for the 2D spatial
case. For the 2D spatiotemporal and the 3D case, we have assumed
that the light bullet experiences anomalous GVD, and has a
dispersion length equal to the diffraction length, i.e. we have
assumed spatiotemporal symmetry for the system, as is evident in
Eq. (\ref{waveeq}). Further examination of Eq. (\ref{critical})
shows that the beam's power is independent of its size for 2D
solitons, which is a well-known property of the Townes solitons,
and the light-bullet's energy decreases as its size decreases in
the 3D case \cite{soliton_collapse4}.

As said above, two different forms of the saturation of the Kerr
nonlinearity were previously considered in detail theoretically,
with $\Delta n(I)$ in rational form
\cite{enns_1,edmundson_1,edmundson_2,blair},
\begin{equation}
\Delta n(I)=\frac{n_{2}I}{\left( 1+I/I_{\mathrm{sat}}\right) }\ ,
\label{twolevelmodel}
\end{equation}
and CQ (cubic-quintic) \cite{quiroga}-\cite{BobPego},
\begin{equation}
\Delta n(I)=n_{2}I-n_{4}I^{2}\ ,  \label{cubicquintic}
\end{equation}
with both $n_{2}$ and $n_{4}$ positive. Although these two models
are usually treated separately (and, as mentioned above, they
produce qualitatively different results for vortex solitons), they
are two approximate forms of the nonlinear index for real
materials. When the light frequency is close to\ a resonance, Eq.
(\ref{twolevelmodel}) describes the system well; if the frequency
is far away from resonance, Eq. (\ref {cubicquintic}) is a better
approximation. When $I\ll I_{\mathrm{sat}}$, Eq.
(\ref{twolevelmodel}) can be expanded, becoming equivalent to the
CQ model,
\begin{equation}
\Delta n(I)\approx n_{2}I-\left( n_{2}/I_{\mathrm{sat}}\right) I^{2}\equiv
n_{2}I-n_{4}I^{2}\ .  \label{twolevelmodelapp}
\end{equation}
with $n_{4}\equiv {n_{2}}/{I_{\mathrm{sat}}}$. The two models produce
essentially different results when the expansion is not valid.

Critical conditions for the formation of 2D solitons in these
systems were found numerically by Quiroga-Teixeiro \textit{et al}.
\cite {quiroga} (2D), and by Edmundson \textit{et al}.
\cite{edmundson_2} and McLeod \textit{et al}. \cite{blair} for the
3D solitons. From those results, we can estimate the necessary
experimental parameters for both the 2D and 3D case by the
transformation to physical units. The transformation is based on
scaling properties of the governing equation (\ref {waveeq}). The
estimate again assumes a Gaussian profile, which yields
\begin{equation}
I\geq I_{\mathrm{stable}}\approx \left\{
\begin{array}{r@{\quad \ \quad}l}
0.16~\left( n_{2}/n_{4}\right)  & \mathrm{for\ \ 2D}, \\
&  \\
1.25\left( n_{2}/n_{4}\right)  & \mathrm{for\ \ 3D},
\end{array}
\right.   \label{stable}
\end{equation}
for the minimum peak intensity needed to launch a stable soliton, and
\begin{equation}
w_{0}\geq w_{\mathrm{stable}}\approx \left\{
\begin{array}{r@{\quad \ \quad}l}
0.77\lambda _{0}\sqrt{n_{0}n_{4}}/n_{2} & \mathrm{for\ \ 2D}, \\
&  \\
0.3\lambda _{0}\sqrt{n_{0}n_{4}}/n_{2} & \mathrm{for\ \ 3D},
\end{array}
\right.   \label{size}
\end{equation}
for the minimum size of the beam. The latter translates into the
minimum diffraction length,
\begin{equation}
L_{\mathrm{diffr}}\geq \left\{
\begin{array}{r@{\quad \ \quad}l}
3.68\lambda _{0}n_{4}\left( n_{0}/n_{2}\right) ^{2} & \mathrm{for\ \ 2D}, \\
&  \\
0.56\lambda _{0}n_{4}\left( n_{0}/n_{2}\right) ^{2} & \mathrm{for\
\ 3D}.
\end{array}
\right.   \label{ldf}
\end{equation}
In the derivation of the above equation, we have used the result
from a CQ model for the 2D case \cite{quiroga}. The validity of
the result can be verified from the fact that
$I_{\mathrm{stable}}(n_4 / n_2) \approx 0.16$, which gives an
error of $(I_{\mathrm{stable}}(n_4 / n_2))^2 \approx 0.025\ll 1$
in the expansion of Eq. (\ref{twolevelmodelapp}). This means it is
appropriate to use a CQ model to determine the boundary where the
solitons start to become stable. On the other hand, the result
from a model with the form of Eq. (\ref{twolevelmodel}) is used
instead for the 3D case \cite{edmundson_2, blair}, which yields a
result of $I_{\mathrm{stable}}\approx 1.25 I_{\mathrm{sat}}$ and
$I_{\mathrm{sat}}$ can always be expressed in $n_{2}$ and $n_{4}$
as described in Eq. (\ref{twolevelmodelapp}).

In general, these results show that the required intensity
decreases with $({ n_{2}}/{n_{4}})$. This means that a larger
self-defocusing coefficient $ n_{4} $ makes it easier to arrest
collapse, as expected. On the other hand, a larger $n_{4}$ also
makes the beam size larger. This is also understandable, since
stronger self-defocusing reduces the overall focusing effect and
makes the beam balanced at a larger size.

\section{Theoretical analysis of necessary conditions for the existence of
two- and three-dimensional solitons: the limitations due to
losses}

Up to this point, the medium was assumed lossless. In real
materials, saturable nonlinear refraction is accounted for by
proximity to a certain resonance, which implies inevitable
presence of considerable loss. Strictly speaking, solitons cannot
exist with the loss. Of course, dissipation is present in any
experiment. The challenge is to build a real physical medium which
is reasonably close to the theoretical models predicting stable
solitons. In particular, this implies, as a goal, the
identification of materials that exhibit the required saturable
nonlinear refraction, with accompanying losses low enough to allow
the observation of the essential features of the solitons. Under
these conditions, only soliton-like beams (``quasi-solitons''),
rather than true solitons, can be produced. Nevertheless, in cases
where losses are low enough for such quasi-solitons to exist (the
conditions will be described below), we refer to the objects as
\textquotedblleft solitons\textquotedblright .

As candidate optical materials for the soliton generation, we
focus on glasses, as they offer a number of attractive properties
\cite {glass1,glass2,glass3}. Their $\chi ^{(3)}$ susceptibility
is, generally, well-known, varying from the value of fused silica
($n_{2}\sim 3\times 10^{-16}$cm$^{2}$/W) up to $1000$ times that
value. The linear and nonlinear susceptibilities of glasses
exhibit an almost universal behavior that depends largely on the
reduced photon energy (${\hbar \omega }/{E_{g}}$, where $\hbar
\omega $ is the photon energy, and $E_{g}$ is the absorption edge,
as defined in Refs. \cite {glass1,glass2,glass3}). This results in
simple and clear trends that can be easily understood. The wide
variety of available glasses offers flexibility in the design of
experiments. Glasses are solid, with uniform isotropic properties
that make them easy to handle and use. There are recent
experimental reports of saturable nonlinearities in some
chalcogenide glasses \cite{bala}. The saturable nonlinearity was
actually measured with the photon energy above the two-photon
absorption edge, hence this case is not relevant to the pulse
propagation, as the loss would be unacceptably high. However,
these measurements encourage the search for situations where the
nonlinearity saturation is appreciable while the loss is
reasonably low.

It is possible to crudely estimate the conditions that will be relevant to
soliton formation based on the general features of the nonlinearities of
glasses. The nonlinearity of the $(2n-1)$th order will become significant
and increase rapidly when the photon energy crosses the $n$-photon
resonance. Just as the nonlinear index increases rapidly (and is accompanied
by two-photon absorption, 2PA) when ${\hbar \omega }/{E_{g}\sim 0.5}$, we
expect $n_{4}$ to become significant (and be accompanied by three-photon
absorption, 3PA) when ${\hbar \omega }/{E_{g}}\sim 0.33$. The requirement
that $n_{4}$ be appreciable without excessive 2PA or 3PA implies that,
within the window $0.33<{\hbar \omega }/{E_{g}}<0.5$, the solitons may be
possible.

To formulate these conditions in a more accurate form, it is necessary to
identify a maximum loss level beyond which the dynamics deviate
significantly from that of a lossless system. This issue can be addressed by
theoretical consideration of quasi-solitons in (weakly) dissipative systems.
First of all, we fix, as a \textit{tolerance limit}, an apparently
reasonable value of $\ell _{\mathrm{tolerance}}\equiv 10\%$
peak-intensity loss per characteristic (diffraction) length,
$L_{\mathrm{diffr}}$. From what follows below, it will be clear
how altering this definition may impact the predicted parameter
window for soliton formation.

If the loss is produced by 2PA, the corresponding evolution equation for the
peak intensity $I(z)$ is
\begin{equation}
\frac{dI}{dz}=-\beta _{\mathrm{2PA}}I^{2},  \label{2PA}
\end{equation}
where $\beta _{\mathrm{2PA}}$ is the 2PA coefficient. It follows
that the loss per $L_{\mathrm{diffr}}$ (provided that the it is
small enough) is $\Delta I\approx -\beta
_{\mathrm{2PA}}I^{2}L_{\mathrm{diffr}}$. The substitution of the
above definition of the tolerance threshold, $|\Delta I|/I<\ell
_{\mathrm{tolerance}}$, into the latter result leads to an upper
bound on the intensity:
\begin{equation}
I<I_{\mathrm{2PA~tolerance}}\equiv \frac{\ell
_{\mathrm{tolerance}}}{\beta _{ \mathrm{2PA}}L_{\mathrm{diffr}}}\
.  \label{losslimit}
\end{equation}
Notice that the condition (\ref{size}) implies that the diffraction length
cannot be too short, hence the upper limit in Eq. (\ref{losslimit}) cannot
be extremely high.

An analogous result for 3PA is
\[
I^{2}<I_{\mathrm{3PA~tolerance}}^{2}\equiv \frac{\ell
_{\mathrm{tolerance}}}{ \beta _{\mathrm{3PA}}L_{\mathrm{diffr}}}\
,
\]
which follows from the evolution equation [cf. Eq. (\ref{2PA})]$
dI/dz=-\beta _{\mathrm{3PA}}I^{3}$. However, as will be discussed
later, in the case relevant to the soliton formation, 2PA
dominates over 3PA.

On the other hand, within the distance necessary for the
observation of the soliton, its peak intensity must remain above
the threshold value (\ref {stable}), to prevent disintegration of
the soliton. Solving Eq. (\ref{2PA}), this sets another constraint
on the intensity:
\begin{equation}
\frac{I_{0}}{1+N\beta
_{\mathrm{2PA}}I_{0}L_{\mathrm{diffr}}}>I_{\mathrm{ stable}}\ ,
\label{observation}
\end{equation}
where $I_{0}$ is the initial peak intensity, and $N$ is the number
of diffraction lengths required for the experiment. In this work,
we assume $N=5 $, which is sufficient for the reliable
identification of the soliton \cite
{quadratic_sts,quadratic_sts2}. Note that the condition
(\ref{observation}) can never be met if the necessary value
$I_{\mathrm{stable}}$ is too high,
\begin{equation}
I_{\mathrm{stable}}>I_{\max }\equiv \left( N\beta_{\mathrm{2PA}}
L_{\mathrm{diffr}}\right) ^{-1}\ .  \label{boundary}
\end{equation}
In the case of $I_{0}\geq I_{\max }$, the overall peak-intensity
loss with the propagation will be $\geq 50\%$. We will refer to
the situation in which $I_{\mathrm{stable}}>I_{\max }$ as a
\textquotedblleft loss dominating\textquotedblright\ one, and the
opposite as \textquotedblleft saturation
dominating\textquotedblright , since $1/I_{ \mathrm{stable}}$ and
$1/I_{\max }$ can be viewed, respectively, as measures of
saturation and loss in the system. When saturation dominates over
the 2PA loss, and hence creation of the soliton is possible, Eq.
(\ref {observation}) can cast into the form of a necessary
condition for the initial peak power,
\begin{equation}
I_{0}>I_{\mathrm{\min }}\equiv
\frac{I_{\mathrm{stable}}}{1-I_{\mathrm{stable }}/I_{\max }}\ .
\label{observation2}
\end{equation}

The material-damage threshold, $I_{\mathrm{damage}}$, also limits
the highest possible peak intensity that can be used
experimentally. Although this threshold depends on both the
material and pulse duration, we will assume
$I_{\mathrm{damage}}\simeq 100\ $GW/cm$^{2}$, which is typical for
nonlinear glasses and pulses with the duration $\sim 100$ fs.
Thus, all the above results can be summarized in the form
\begin{equation}
I_{\mathrm{\min }}<I_{0}<\min \left\{
I_{\mathrm{2PA~tolerance}},I_{\mathrm{ damage}}\right\} .
\label{window}
\end{equation}
In a material with known nonlinearity and loss, experimental observation of
the solitons is feasible if the corresponding window (\ref{window}) exists.

A somewhat simplified but convenient way to assess this is to
define a figure of merit (FOM). In the case when
$I_{\mathrm{damage}}>I_{\mathrm{ 2PA~tolerance}}$,
\begin{eqnarray}
\mathrm{FOM} &\equiv &\log \left(
\frac{I_{\mathrm{2PA~tolerance}}}{I_{
\mathrm{\min }}}\right)   \nonumber \\
&=&\left\{
\begin{array}{r@{\quad \quad}l}
\log \left[ \ell _{\mathrm{tolerance}}\left( 1.74\frac{n_{2}}{
n_{0}^{2}\lambda _{0}\beta _{\mathrm{2PA}}}-N\right) \right] \ , &
\\
\\
 \mathrm{
for\ 2D}, &\\
\\
\\
\log \left[ \ell _{\mathrm{tolerance}}\left( 1.42\frac{n_{2}}{
n_{0}^{2}\lambda _{0}\beta _{\mathrm{2PA}}}-N\right) \right] , &
\\
\\
\mathrm{ for\ 3D}.&
\\
\\
\end{array}
\right.   \label{fom1}
\end{eqnarray}
If $I_{\mathrm{damage}}$ is smaller than $I_{\mathrm{2PA~tolerance}}$, the
definition becomes
\begin{eqnarray}
\mathrm{FOM} &\equiv &\log \left(
\frac{I_{\mathrm{damage}}}{I_{\mathrm{\min }
}}\right)   \nonumber \\
&=&\left\{
\begin{array}{r@{\quad \quad}l}
\log \left[ I_{\mathrm{damage}}\frac{n_{4}}{n_{2}}\left(
6.4-3.68N\beta_{\mathrm{2PA}} \frac{ \lambda
_{0}n_{0}^{2}}{n_{2}}\right) \right] \ , &
\\
\\ \mathrm{for\ 2D}, &
\\
\\
\\
\log \left[ I_{\mathrm{damage}}\frac{n_{4}}{n_{2}}\left(
0.8-0.56N\beta_{\mathrm{2PA}} \frac{ \lambda
_{0}n_{0}^{2}}{n_{2}}\right) \right] \ , &
\\
\\ \mathrm{for\ 3D}. &
\\
\\
\end{array}
\right.   \label{fom2}
\end{eqnarray}
The FOM is a measure of the range between the minimum required and maximum
allowed values of the peak intensity. Of course, it must be positive, and
the larger the FOM, the better the chance to observe solitons.

It seems to be commonly accepted that a larger quintic
self-defocusing coefficient $n_{4}$ is always desirable, but the
above results show that this is not always true. From the FOM we
can see that a larger $n_{4}$ is better in the sense that it
reduces the lower threshold $I_{\min }$, helping to secure the
positiveness of the FOM (\ref{fom2}). However, as soon as $
I_{\min }$ is low enough that the damage threshold no longer poses
a problem, Eq. (\ref{fom1}) shows that larger $n_{4}$ does not
help, and the loss factor $\beta _{\mathrm{2PA}}$ dominates. One
can understand this, noticing that, although larger $n_{4}$
reduces $I_{\min }$, at the same time it increases the beam's
width and makes the needed experimental propagation length longer,
as is clearly shown by Eq. (\ref{ldf}). In turn, more loss
accumulates due to a longer propagation length, which offsets the
benefit of a lower $I_{\min }$.

\section{Measurements of nonlinear parameters of glasses}

The eventual objective is to answer the following question: for a
given category of materials (such as glasses), with known
nonlinear, loss, and damage characteristics, does there exist a
combination of material and wavelength such that solitons can be
observed? To this end, we have measured the nonlinearity in a
series of glasses with $100$-fs pulses from a Ti:sapphire
regenerative amplifier with center wavelength at $790\ $nm.
Sapphire is used (it has ${\hbar \omega } /{E_{g}}\cong 0.25$ in
this case) as a reference material with minimum nonlinearity.
Although fused silica can also be used for this purpose,
sapphire's higher damage threshold allows us to measure at higher
intensities.

We measured several glasses, including: SF$59$ (with ${\hbar
\omega }/{E_{g}} \simeq 0.5$), La-Ga-S(with ${\hbar \omega
}/{E_{g}}\simeq 0.56$), and As$_{2} $S$_{3}$ (with ${\hbar \omega
}/{E_{g}}\simeq 0.75$). To determine the effective $\chi ^{(3)}$
and $\chi ^{(5)}$ susceptibilities, spectrally resolved two-beam
coupling (SRTBC) was used \cite{srtbc}. We extended the
application of this method to take into account both higher-order
nonlinearities and strong signals \cite{prep}. In general, 2PA is
observable even for ${\hbar \omega }/{E_{g}}<0.5$ owing to the
absorption-edge broadening present in all glasses.

\begin{figure}[th]
\centerline{\scalebox{1}{\includegraphics{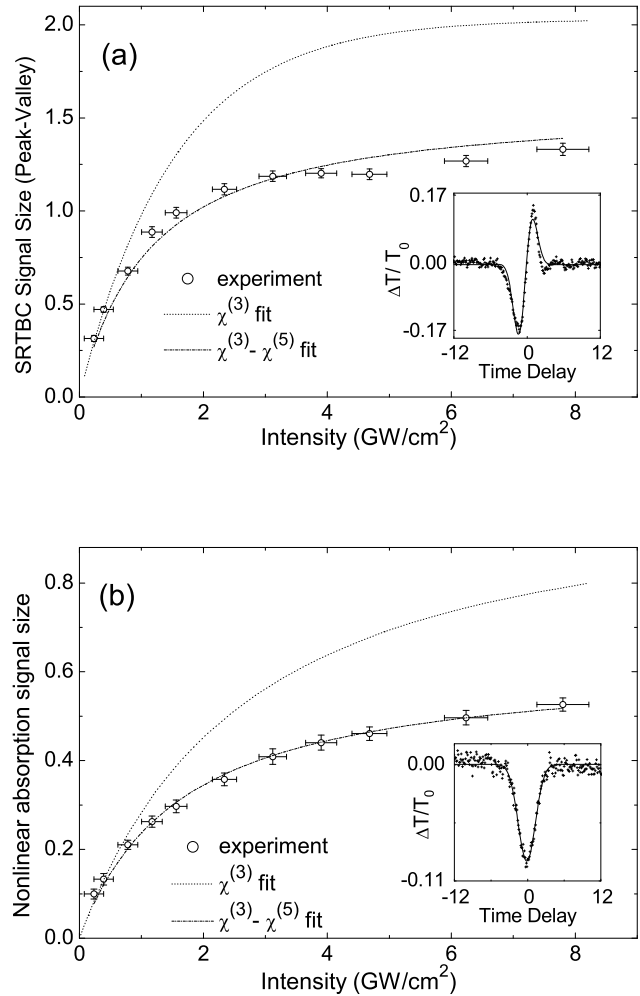}}}
\caption{Intensity dependence of (a) SRTBC signal magnitude
(normalized peak-valley transmission difference) and (b) nonlinear
absorption signal of As$_{2}$S$_{3}$. The saturation of the cubic
nonlinearity is evident. Higher-order nonlinearities, such as
$\protect\chi ^{(5)}$, can be estimated from the deviation from
$\protect\chi ^{(3)}$. Insets show examples of SRTBC and nonlinear
absorption traces (symbols) along with the best-fit theoretical
curves (solid lines). The time delay (on the horizontal axis) is
given in units of the pulse duration (FWHM).} \label{srtbcdata}
\end{figure}

Typical experimental traces obtained from As$_{2}$S$_{3}$ are shown in the
insets of Fig. \ref{srtbcdata}, along with the theoretical fits. The
intensity dependence of SRTBC signal magnitude and normalized nonlinear
absorption signal magnitude are shown in Fig. \ref{srtbcdata}.
The dotted curves in both panels are predictions for the pure
$\chi ^{(3)}$ nonlinearity. The deviation of the experimental
points from these curves evidences the saturation of the
nonlinearity. Postulating the presence of the $\chi ^{(5)}$
self-defocusing nonlinearity provides for good agreement with the
experiments. Similar results were produced by all four samples
used in the measurements; in particular, in all the cases the sign
of the real part of $\chi ^{(5)}$ turns out to be opposite to that
of $\chi ^{(3)}$, i.e., the quintic nonlinearity is
self-defocusing indeed. The measured $\chi ^{(3)}$ coefficients
are consistent with previously reported values \cite {bala,
glassmeasurement_1,glass1}.

From these results, we also observe that higher-order
nonlinearities become more important as the optical frequency
approaches a resonance, as expected on physical grounds. The $\chi
^{(5)}$ part of the nonlinearity is most significant for
As$_{2}$S$_{3}$, while for sapphire it is below the detection
threshold.

\section{Stability windows for the two- and three-dimensional solitons}

\begin{figure}[th]
\centerline{\scalebox{1}{\includegraphics{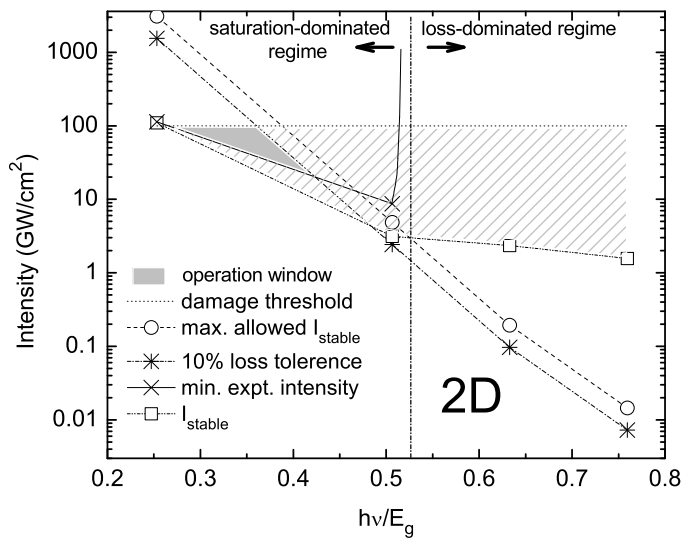}}} \caption{The
operation window for the 2D solitons, as predicted on the basis of
the experimentally-measured characteristics of the glass. The
hatched area is the window neglecting the loss. The shaded area is
the dramatically reduced (but definitely existing) window found
with the loss taken into account.} \label{multipanel2d}
\end{figure}

The measurements provide the information needed to construct the
window for the soliton formation. The results for 2D case are
shown graphically in Fig. \ref{multipanel2d}.
The intensity limitations are plotted on the diagram against the
reduced photon energy. The parameter space can be divided into two
regions which were defined above, viz., the saturation-dominating
and absorption-dominating ones, with the boundary between then
determined by Eq. (\ref{boundary}). To demonstrate the dramatic
effect of the loss, we also plot the window for the (unrealistic)
case when loss is completely neglected (the hatched area). In the
absence of loss, the window is very large and the FOM increases
monotonically with the reduced photon energy. The shaded area is
the window remaining after inclusion of the loss. It is greatly
reduced compared to the lossless case, and the best FOM is
obtained near ${\hbar \omega }/{E_{g}}\simeq 0.35$. From this
diagram, we conclude that, while the saturation of the
nonlinearity is definitely necessary to stabilize the soliton,
major restrictions on the window are imposed by the loss.

From the above rough estimation that were based on the band-edge arguments,
one might expect that 3PA would further curtail the window, when the 2PA
effects are weak (which is the case exactly inside the predicted window).
However, $n_{2}$ and 2PA have been observed in glasses for the reduced
photon energy as low as $\sim 0.35$ \cite{glass2}, due to the fact that the
band edge in glasses extends well below the nominal value. Since significant
2PA remains in this region, 3PA may be neglected indeed. Hence, 2PA presents
the fundamental limitation to observing solitons in these media [as
quantified by the FOM in Eqs. (\ref{fom1}) and (\ref{fom2})].

\begin{figure}[th]
\centerline{\scalebox{1}{\includegraphics{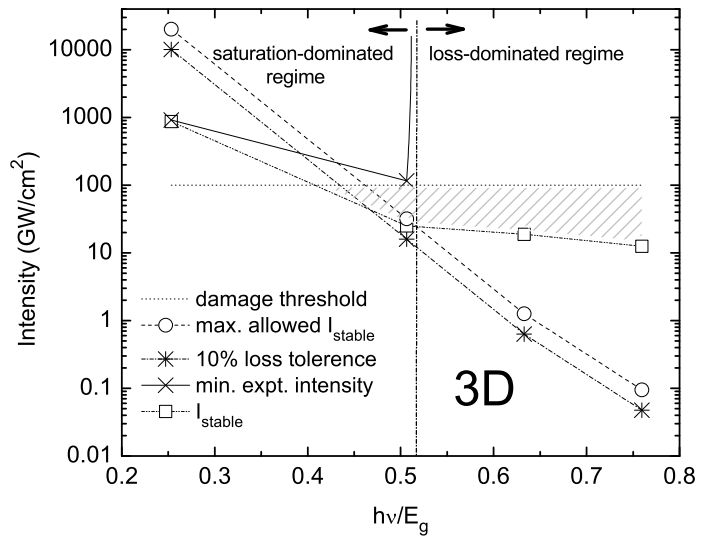}}} \caption{The
operation window for 3D solitons. The meaning of the hatched area
is the same as in the 2D case, i.e., it shows the window obtained
neglecting the loss. When the loss is taken into account, the
window vanishes completely.} \label{multipanel3d}
\end{figure}

The results of the analysis for the 3D solitons are summarized in
Fig. \ref {multipanel3d}. Note that another major issue in this
case is the requirement of anomalous GVD. This requirement is
neglected here (addition of it will only further constrain the
window, which does not really exist even without that, see below).
From Fig. \ref{multipanel3d}, we observe that, even in the
lossless case, the window (hatched area) is significantly smaller
than in the 2D case. This is expected, because collapse is
stronger in 3D than 2D \cite{Luc}. As in the 2D case, the loss
again is a major concern for performing experiments. The most
important inference is that the window closes up completely when
loss is taken into account. Thus, it appears that loss will
preclude the creation of 3D solitons in glasses, while leaving
room for the 2D solitons.

Our overall conclusion is that a challenge is to perform
experimental studies of 2D solitons in saturable Kerr media. Both
spatial and spatiotemporal solitons are possible to be produced
experimentally. Among these two, the 2D spatiotemporal case is
more complicated since it requires anomalous GVD. In general, this
will naturally constrain the window further. On the other hand, in
this case tilted-pulse techniques could be used to obtain
anomalous GVD. It is also possible to use a planar waveguide to
perform 2D spatiotemporal soliton experiments.

Of course, the predicted window depends on the assumed parameters
(such as the damage threshold) and criteria (such as the $10\%$
loss per diffraction length).Variations in these parameters will
naturally impact the window, and our analysis provides the
guidelines for searching for the most favorable materials and
wavelength. A next natural step is to perform numerical
simulations of the pulse propagation with the parameters selected
in the present work. It is conceivable that the window for 3D
solitons would finally open through variations of material
parameters. In that case, one would still have to find an overlap
of the resulting window with the condition that the GVD must be
anomalous. More generally, non-glass materials may be tried to
improve the possibilities for the experiment.

\section{Conclusion}

We have developed criteria for experimental observation of
multi-dimensional solitons -- spatial and spatiotemporal 2D
solitons and spatiotemporal 3D ones. Using these criteria and
measured properties of nonlinear glasses within a range of reduced
photon energies, we have shown that the loss that accompanies
higher-order nonlinearities (which are tantamount to saturation of
the cubic nonlinearity) will set very stringent limits on the
material parameters appropriate for the experiment. While loss was
thus far neglected in theoretical treatments of multi-dimensional
solitons, this work motivates more systematic
studies of the soliton-like propagation in lossy media. \\
\indent The criteria developed in this paper can also be applied,
as an assessment tool, to materials other that glasses. More
generally, the same rationale used for obtaining the relevant
boundaries in this paper can also be used in systems other than
optical ones. In these cases the specific mathematical forms of
the boundaries will be different. In any case, the analysis
presented here suggests that there is a small but apparently
usable window of parameters in which 2D solitons can be generated,
and work is underway to address this possibility. On the other
hand, the prospects for generating 3D solitons in glasses are
quite poor.
\begin{acknowledgments}
This work was supported by the National Science Foundation under grant
PHY-0099564, and the Binational (U.S.-Israel) Science Foundation (Contract
No. 1999459). We thank Jeffrey Harbold for valuable discussions.
\end{acknowledgments}

\end{document}